\documentclass[onecolumn,%
11pt,tightenlines,
notitlepage,%
nofootinbib,%
superscriptaddress,%
amsfonts,amsmath]{revtex4}

\begin{document}

\newcommand{\beq} {\begin{equation}}
\newcommand{\eeq} {\end{equation}}
\newcommand{\bea} {\begin{eqnarray}}
\newcommand{\eea} {\end{eqnarray}}
\def\GB{{\hat{\cal{G}}}}
\newcommand{\RR}{(*R*)}
\def\third{\textstyle{1\over3}}
\def\quarter{\textstyle{1\over4}}
\newcommand{\G}{\mathcal G}
\title{Beyond Fab Four}

\author{E.~Babichev} 
\affiliation{Laboratoire de Physique Th\'eorique d'Orsay,
B\^atiment 210, Universit\'e Paris-Sud 11,
F-91405 Orsay Cedex, France}

\author{C.~Charmousis} 
\affiliation{Laboratoire de Physique Th\'eorique d'Orsay,
B\^atiment 210, Universit\'e Paris-Sud 11,
F-91405 Orsay Cedex, France}

\author{D. Langlois}
\affiliation{%
APC (CNRS-Universite Paris 7), 10 rue Alice Domon et L\'eonie Duquet, 75205 Paris, France
}%

\author{R. Saito}
\affiliation{%
APC (CNRS-Universite Paris 7), 10 rue Alice Domon et L\'eonie Duquet, 75205 Paris, France
}%

\begin{abstract}
We show that the two additional Lagrangians that appear in theories beyond Horndeski  can be reexpressed in terms of simple generalizations of the ``John'' and ``Paul'' terms of the Fab Four theories.  We find that these extended Fab Four satisfy the same  properties of self-tuning as the original Fab Four.
\end{abstract}

\date{\today}


\maketitle

\section{Introduction}

As variations on the theme of modified gravity, scalar tensor theories have often played a central role. Recently, a lot of attention has been given to Horndeski's theory, the most general scalar tensor theory leading to second order equations of motion, published in a 1974 paper~\cite{horndeski} that remained little known until resurrected in \cite{Charmousis:2011bf}, and rediscovered in \cite{Deffayet:2011gz} in the guise of so-called generalized galileons. 
\footnote{
The generalized galileons have been shown to be equivalent to Horndeski's theories in \cite{Kobayashi:2011nu}.
}
Horndeski's action, which contains second order derivatives of the scalar field, encompasses most of the previously studied scalar tensor theories and was long believed to be the most general consistent scalar-tensor theory in the sense that it does not suffer from ghost-like instabilities that usually plague theories with higher order spacetime gradients. However, it was recently realized  that  Horndeski's Lagrangian can be extended without encountering unwanted extra degrees of freedom~\cite{Gleyzes:2014dya,Gleyzes:2014qga} 
(see also \cite{Lin:2014jga,Deffayet:2015qwa, Langlois:2015cwa} for complementary analyses and \cite{Bettoni:2015wta} for another approach to construct theories beyond Horndeski). 
These extensions also lead to interesting new phenomenology, such as a partial  breaking  of the Vainshtein mechanism inside matter~\cite{Kobayashi:2014ida,Koyama:2015oma,Saito:2015fza} or a potentially different behavior of linear cosmological perturbations~\cite{Gleyzes:2013ooa,Gleyzes:2014rba}.

As shown in \cite{Charmousis:2011bf, Charmousis:2011ea}, Horndeski's theory contains  a sub-class of  Lagrangians that enjoy the very special property of self-tuning. The idea of self-tuning, developed notably in the braneworld paradigm (see e.g. \cite{selftuning}), relies on some mechanism that enables to hide an arbitrarily large cosmological constant, i.e. which admits a Minkowski spacetime solution in presence of a cosmological constant.  In the context of scalar tensor theories, it requires the existence of a Minkowski spacetime solution with a non-trivial scalar in the presence of an arbitrary cosmological constant. 
As demonstrated in \cite{Charmousis:2011bf, Charmousis:2011ea}, the self-tuning theories within Horndeski, known as  the Fab Four, consist of combinations of the following four Lagrangians that depend on a scalar field $\phi$ and a metric $g_{\mu\nu}$: 
\bea
{\cal L}_{john} &=& \sqrt{-g}\, V_J(\phi)\, G^{\mu\nu} \, \nabla_\mu\phi \nabla_\nu \phi \,,
\\
{\cal L}_{paul} &=&\sqrt{-g} \, V_P(\phi) \,   P^{\mu\nu\alpha \beta} \, \nabla_\mu \phi \nabla_\alpha \phi \nabla_\nu \nabla_\beta \phi
\,,
  \\
{\cal L}_{george} &=&\sqrt{-g}\, V_G(\phi)\,  R 
\,,
\\
{\cal L}_{ringo} &=& \sqrt{-g}\, V_R(\phi) \, \hat G\,,
\eea
where $G_{\mu\nu}$ is the Einstein tensor, $P^{\mu\nu\alpha \beta}$ the double dual of the Riemann tensor, $R$ the scalar curvature and  $\hat G$ the Gauss-Bonnet term.

Given the recent extension beyond Horndeski discussed above, it is natural to wonder whether the subspace of self-tuning theories can be extended in this larger space of theories, in other words whether one can find new theories, beyond the Fab Four, that also satisfy this self-tuning property. The goal of this paper is to show that the answer is indeed positive. Moreover, we find that the two Lagrangians beyond Horndeski can be rewritten, up to Horndeski Lagrangians,  as straightforward generalizations of 
the John and Paul terms where the arbitrary functions $V_J(\phi)$ and $V_P(\phi)$ are simply replaced by arbitrary functions of $\phi$ and of the standard kinetic term $X\equiv\nabla_\mu\phi\nabla^\mu\phi$. These extended John and Paul satisfy the properties of self-tuning, as we show explicitly by writing down their contributions to the cosmological equations of motion, assuming spacetime homogeneity and isotropy. 
Remarkably, the stealth black hole solutions discovered in \cite{Babichev:2013cya} remain valid with the extended John Lagrangian.

Our plan is the following. In the next section, we show explicitly how the extended John and Paul Lagrangians can be expressed as combinations of Horndeski Lagrangians and the terms beyond Horndeski presented in \cite{Gleyzes:2014dya}. We then write the corresponding equations of motion. In the subsequent section, we derive the cosmological equations for the extended Fab Four and show that they satisfy the same self-tuning properties as the original Fab Four.  We conclude in the final section. In an appendix, we also  give the ADM formulation of the extended John and Paul terms in the uniform scalar field gauge.

\section{Beyond Horndeski Lagrangian and equations of motion}\label{sec:BF4}

In this section,  we show that the extended John and Paul Lagrangians where the functions $V_J$ and $V_P$ are replaced by functions of $\phi$ and $X$ can be rewritten as Lagrangians beyond Horndeski combined with standard Horndeski terms.
We thus consider the following two actions,
\begin{eqnarray}
	S_{J}^{\rm ext} &=& \int\ d^4x\ \sqrt{-g}\  F_J(\phi,X)\ G^{\mu\nu} \nabla_\mu\phi\nabla_\nu\phi \,,\label{bH4}\\
	S_{P}^{\rm ext} &=& \int\ d^4x\ \sqrt{-g}\  F_P(\phi,X)\ P^{\alpha\beta\mu\nu} \nabla_\alpha\phi\nabla_\mu\phi\nabla_\beta\nabla_\nu\phi \,,\label{bH5}
\end{eqnarray}
where $F_J$ and $F_P$ are two arbitrary functions of $\phi$ and  $X$. In the above expressions,  
$G^{\mu\nu}$ is the Einstein tensor 
and $P^{\alpha\beta\mu\nu}$  is the double dual of the Riemann tensor, 
	\begin{align}
		P_{\alpha\beta\mu\nu} & \equiv -\frac14 \epsilon_{\alpha\beta\rho\sigma} R^{\rho\sigma\gamma\delta} \epsilon_{\mu\nu\gamma\delta} \nonumber \\
			&= R_{\alpha\beta\mu\nu}+ g_{\alpha\nu} R_{\beta\mu} - g_{\alpha\mu} R_{\beta\nu} + g_{\beta\mu} R_{\alpha\nu}-g_{\beta\nu} R_{\alpha\mu} 
			+\frac12 \left( g_{\alpha\mu}g_{\beta\nu} - g_{\alpha\nu}g_{\beta\mu}\right) R\,,
	\end{align}
where $\epsilon_{\alpha\beta\mu\nu}$ is the  totally antisymmetric Levi-Civita tensor.

Using the Ricci identity and making use of integration by parts, one can derive from the action (\ref{bH4}), up to total derivative terms that are irrelevant, the new expression
\begin{equation}\label{bH4bis}
	\begin{aligned}
	S_{J}^{\rm ext} =&  \int\ d^4x\ \sqrt{-g}\  \left\{  
	F_{J,X}\ \epsilon_{\mu\gamma\alpha\beta}\ \epsilon_{\nu\delta\rho}^{\phantom{\nu\delta\rho}\beta}\ \phi^{;\mu}\phi^{;\nu}\phi^{;\delta\gamma}\phi^{;\alpha\rho}  \right .\\
	&\left.  + \left(F_JX\right)_{,X}\left[\left(\Box\phi\right)^2 - \left(\nabla\nabla\phi\right)^2\right] 
	 - \frac12 X F_J R  +F_{J,\phi}\left( X\Box\phi -\phi_{;\alpha}\phi_{;\beta}\phi^{;\alpha\beta} \right) \right\},
	\end{aligned}
\end{equation}
where we use the notations $F_{,X}\equiv \partial F/(\partial X)$ and $F_{,\phi}\equiv \partial F/(\partial \phi)$ and $\left(\nabla\nabla\phi\right)^2 \equiv \phi_{;\alpha\beta}\phi^{;\alpha\beta}$.
Notice that the second line of (\ref{bH4bis}) belongs to Horndeski Lagrangians, as it is a sum of two generalized Galileon Lagrangians.
The first term inside the integral of (\ref{bH4bis}) coincides with the beyond Horndeski term given in \cite{Gleyzes:2014dya}, 
with the identification $F_{J,X\, [here]} = F_{4\, [there]}$. 
Therefore, up to the Horndeski terms, Eq.~(\ref{bH4}) corresponds to one of the two beyond Horndeski terms presented in~\cite{Gleyzes:2014dya}. 

Similarly, using the Ricci identity, integrating by parts (\ref{bH5}) and using (\ref{bH4bis}), one 
 finds
\begin{equation}\label{bH5bis}
	\begin{aligned}
	S^{\rm ext}_{P} =&  \int\ d^4x\ \sqrt{-g}\  \left\{  
	- \frac{F_{P,X}}{3}\ \epsilon_{\mu\gamma\alpha\beta}\ \epsilon_{\nu\delta\rho\sigma}\ \phi^{;\mu}\phi^{;\nu}\phi^{;\delta\gamma}\phi^{;\alpha\rho}\phi^{;\beta\sigma}  \right .\\
	& -  \frac13 \tilde F_{P,X}\left[\left(\Box\phi\right)^3 - 3 (\Box\phi)\left(\nabla\nabla\phi\right)^2 +2 \left(\nabla\nabla\phi\right)^3 \right] 
	 - \tilde F_P G^{\mu\nu}\phi_{;\mu\nu} \\
	 &\left.   -\frac12 \left(\tilde C_P X\right)_{,X}\left[\left(\Box\phi\right)^2 - \left(\nabla\nabla\phi\right)^2\right] 
	 + \frac14 X \tilde C_P R  -\frac12 \tilde P_{P,\phi}\left( X\Box\phi -\phi_{;\alpha}\phi_{;\beta}\phi^{;\alpha\beta} \right) \right\},
	\end{aligned}
\end{equation}
where we have introduced the functions
$$
\tilde F_P \equiv  F_P X +\frac12 \int F_P dX \,,\quad \tilde C_P \equiv \int F_{P, \phi} dX \,.
$$
One recognizes in the last two lines  of (\ref{bH5bis}) terms that belong to Horndenski class, while the first line coincides with the second beyond Horndeski term given in \cite{Gleyzes:2014dya},
with identification $F_{P,X\, [here]} = -3 F_{5\, [there]}$.
 
As shown in \cite{Gleyzes:2014dya,Gleyzes:2014qga}, the Lagrangians 
within Horndeski and beyond Horndeski 
possess a remarkably simple form when written in ADM form in the gauge where the scalar field is uniform, also called unitary gauge. In this gauge, the Lagrangian depends only 
on the lapse and on the intrinsic and extrinsic curvature tensors, without time derivative that could signal the presence of a ghost-like degree of freedom. 
In the appendix, we provide the corresponding ADM formulation for the extended John and Paul Lagrangians. 
We also show that, by contrast, the George and Ringo terms cannot be generalized into ADM Lagrangians with similar properties.
 
 The field equations for the extended Fab Four are obtained from the variation of the actions (\ref{bH4}) and (\ref{bH5}), in place of John and Paul,  while George and Ringo are unchanged\footnote{See \cite{Charmousis:2011ea} for the full field equations of the original Fab Four. Note that we defined the double dual of the Riemann tensor with the opposite sign of theirs.}.
 The contribution of the extended John and Paul Lagrangians to the metric field equations is of the form  ${\cal E}^{\mu\nu} \equiv -\frac{1}{\sqrt{-g}}\frac{\delta S}{\delta g_{\mu\nu}}$, with respectively
 	\begin{align}
	\label{eomJg}
		{\cal E}^{\mu\nu}_J =  \frac{1}{2}(XF_JG^{\mu\nu}+ 2 F_J P^{\mu\alpha\nu\beta}\phi_{;\alpha}\phi_{;\beta})
		- \frac{1}{2}\epsilon^{\mu\alpha\sigma\gamma} \epsilon_{\phantom{\nu\beta\rho}\gamma}^{\nu\beta\rho}
		\left( F_J \phi_{;\alpha\beta}\phi_{;\rho\sigma} +2 F_{J;(\alpha}\phi_{;\beta)} \phi_{;\rho\sigma} -\phi_{;\alpha}\phi_{;\beta} F_{J;\rho\sigma}\right) \nonumber\\
		+ F_{J,X} \phi^{;\mu}\phi^{;\nu}G^{\alpha\beta}\phi_{;\alpha}\phi_{;\beta} \,,
	\end{align}
and
	\begin{align}
		&{\cal E}^{\mu\nu}_P = - \frac{1}{2}P^{\mu\alpha\nu\beta}\left( 3XF_P \phi_{;\alpha\beta} + 2XF_{P;(\alpha}\phi_{;\beta)} - F_{P;\lambda}\phi^{;\lambda}\phi_{;\alpha}\phi_{;\beta} \right) \nonumber \\
		&\quad - \frac{1}{2}\epsilon^{\mu\alpha\sigma\gamma}\epsilon^{\nu\beta\rho\tau} \left( F_P \phi_{;\alpha\beta}\phi_{;\rho\sigma}  + 2F_{P;(\alpha}\phi_{;\beta)}\phi_{;\rho\sigma} - \phi_{;\alpha}\phi_{;\beta}F_{P;\sigma\rho} \right)\phi_{;\gamma\tau} 
		- F_{P,X} \phi^{;\mu}\phi^{;\nu}P^{\alpha\beta\rho\sigma} \phi_{;\alpha}\phi_{;\rho} \phi_{;\beta\sigma} \,.
	\end{align}
On the other hand, the scalar field equations, ${\cal E}^{\phi} \equiv -\frac{1}{\sqrt{-g}}\frac{\delta S}{\delta \phi}=0$
become
	\begin{equation}\label{eomphi}
		{\cal E}^{\phi}_J = 2\nabla_\mu\left(F_J G^{\mu\nu}\phi_{;\nu}\right) + 2 \nabla_\alpha\left( F_{J, X} G^{\mu\nu}\phi_{;\mu}\phi_{;\nu}\phi^{;\alpha}\right) - F_{J,\phi} G^{\mu\nu} \phi_{;\mu} \phi_{;\nu} \,,
	\end{equation}
and
	\begin{align}
	\label{eomphi2}
		{\cal E}^{\phi}_P = 2\nabla_{\alpha}(F_PP^{\alpha\beta\mu\nu} \phi_{;\mu}\phi_{;\beta\nu}) - \nabla_{\nu}\nabla_{\beta}(F_P P^{\alpha\beta\mu\nu}\phi_{;\alpha}\phi_{;\mu}) + 2\nabla_{\rho}(F_{P,X}P^{\alpha\beta\mu\nu}\phi_{;\alpha}\phi_{;\mu}\phi_{;\beta\nu}\phi^{;\rho}) & \nonumber \\
		- F_{P,\phi}P^{\alpha\beta\mu\nu}\phi_{;\alpha}\phi_{;\mu}\phi_{;\beta\nu} \,,
	\end{align}
respectively.
When $F_J$ and $F_P$ do not depend on $X$, one recovers the standard Fab Four equations of motion.

\section{Self tuning}

The aim of self-tuning is to hide classically{\footnote{The self tuning paradigm does not take into account radiative corrections that will appear, around a self tuning vacuum, at the ultra violet cut off scale of any classical theory and may destabilize the self tuning mechanism for the cosmological constant. For recent progress on that front the interested reader should consult the recent works \cite{padilla}.}} a large cosmological constant. In the context of scalar tensor theories, self-tuning corresponds to  the existence of a Minkowski spacetime solution with a non-trivial scalar field in the presence of an arbitrary cosmological constant. In other words,  the metric is insensitive to a big cosmological constant which is ``screened'' by a scalar field,  without fine tuning any of the coupling constants of the theory. This means that  a self-tuning solution must fix the cosmological constant dynamically via an integration constant associated to the scalar field and not from the parameters of the theory. As such, vacuum energy would not gravitate, unlike other forms of matter do, breaking explicitly the strong equivalence principle.
 
 By studying Horndeski theory on FLRW backgrounds,~\cite{Charmousis:2011bf} identified  a subclass of models\footnote{We suppose "minimal coupling" of Horndeski to matter i.e. that any matter fields do not couple directly to the scalar field and only to the metric.}, dubbed the Fab Four, leading to a viable self-tuning mechanism. Several criteria were required in order to obtain a  viable self-tuning:
\begin{itemize}
\item The self tuning theory admits a Minkowski spacetime solution for any value of the cosmological constant. 

\item 
This must remain true even if the the cosmological constant jumps abruptly to another value during the time evolution of the Universe. In other words, the self tuning solution must allow for sources with Heaviside distributions, which would mimic some phase transition in the Early Universe.
\item Matter other than vacuum energy does gravitate and sources the spacetime metric.
\end{itemize}
Within Horndeski class, this self-tuning filter leads to a severe restriction since only the Fab Four, characterized by four arbitrary functions of $\phi$, survive among theories that depend on  four arbitrary functions of $\phi$ and $X$. Given the simple way we have rewritten the Beyond Horndeski Lagrangians (\ref{bH4})-(\ref{bH5}), a natural  question is whether these extensions are also of the self-tuning type. 

In order to answer this question we follow the strategy adopted in ~\cite{Charmousis:2011bf} and write the equations of motion in  FLRW cosmology, with metric 
\beq
g_{\mu\nu}dx^\mu dx^\nu=-dt^2 +a^2(t)\left[\frac{dr^2}{1-\kappa r^2}+r^2(d\theta^2+\sin^2\theta d\phi^2) \right] \,, 
\eeq
where $\kappa=1,-1,0$ is the normalized spatial curvature. 

We consider  the total action
\beq
S_{\rm total}=S^{\rm ext}_J+S^{\rm ext}_P+ S_G+S_R -\int d^4 x \sqrt{-g}\ \Lambda \,,
\eeq
where $S_G$ and $S_R$ are the standard George and Ringo actions. One can derive the cosmological equations of motion either by using the general equations of motion specialized to FLRW or by varying the minisuperspace version of the action. One finds that the first Friedmann equation is given by  
\beq
{\cal H}^{\rm ext}_{J}+{\cal H}^{\rm ext}_{P}+{\cal H}_{G}+{\cal H}_{R}=-2 \Lambda \,,
\eeq
with
\bea
{\cal H}^{\rm ext}_{J}&=&{3\dot\phi^2 \left(F_J+2 X F_{J,X}  \right)\left(H^2+\frac{\kappa}{a^2}\right)+}6 \dot\phi^2 H^2 F_J \,, \\
{\cal H}^{\rm ext}_{P}&=&{3 H \dot{\phi}^3 \left(3F_P+2 X F_{P,X}  \right)\left(H^2+\frac{\kappa}{a^2}\right)+}6 \dot\phi^3H^3 F_P \,, \\
{\cal H}_{G}&=&-6V_{G}(\phi){\left(H^2+\frac{\kappa}{a^2}\right)-} 6H\dot\phi V'_G \,, \qquad\\
{\cal H}_{R}&=&{-24\dot\phi H\left(H^2+\frac{\kappa}{a^2}\right)}V'_{R}(\phi) \,.
\eea
As one can see, the extended John and Paul actions simply  lead to an additional  term of the form $2X F_{,X}$ in the contributions, with respect to the standard case where $F_J$ and $F_P$ are independent of $X$.

Let us now consider Minkowski spacetime solutions, which satisfy  the condition 
\beq
\label{flat}
H^2+\frac{\kappa}{a^2}=0\,.
\eeq
This includes the trivial case $\kappa=0$ and $a=$const, but also the Milne solution 
  \beq 
ds^2=-dt^2+t^2\left(\frac{dr^2}{1+ r^2} + r^2 d \Omega^2 \right)\,,
\eeq
which corresponds to a slicing of Minkowski with negative curvature ($\kappa=-1$) spatial hypersurfaces. When the condition (\ref{flat}) is imposed on the Friedmann equation, the new contributions disappear and one recovers exactly the Fab Four situation, up to the fact that $F_J$ and $F_P$ also depend on $\dot\phi$. The Friedmann equation then reads 
\beq
F_J(\phi,X)\dot\phi^2 H^2+F_P(\phi,X)\dot\phi^3H^3- H\dot\phi V'_G(\phi)+\frac{\Lambda}{3}=0\,.
\eeq
What is essential for self-tuning is that  $\dot\phi$  does not drop out 
of the Friedmann equation. Indeed, an abrupt shift of the cosmological constant  value (which  can be approximated  by a Heaviside distribution) can then be compensated by a discontinuity of $\dot\phi$, while $\phi$ is assumed to be continuous. 

One can also derive the equation of motion for the homogeneous scalar field. It is immediate to check that this equation of motion vanishes identically when the Minkowski spacetime condition (\ref{flat}) is imposed, since (\ref{eomphi}) and (\ref{eomphi2}) automatically vanish when the curvature tensors are zero.
This property is in fact necessary  to obtain a viable self-tuning. Otherwise, since the vacuum energy source does not appear in the scalar field equation of motion, the Dirac distributional terms due to the presence of $\ddot \phi$ in the equation would have to be set to zero with $\dot\phi$ being continuous.  This explains the curvature couplings in the Fab Four action and their extension. In order to satisfy the third requirement, one must also check that the scalar field equation is dynamical when (\ref{flat}) is relaxed, 
 thus permitting non-trivial cosmology for other types of matter such as radiation or dust. Indeed, the scalar field should not self tune other forms of matter other than vacuum energy !

We  see that in order for self tuning to work we require, that either $F_J$, or $F_P$ are non zero or again  $V_G$ is non constant. In particular,  GR does not self-tune and, although  the Gauss-Bonnet term does not spoil self-tuning, it needs the presence of any of the other Fab Four terms in order to have a self-tuning solution. 

To see a simple and explicit solution we consider the simplest of potentials corresponding to a slowly varying scalar field for late time behavior,
\beq
F_J = C_j \,,  F_P=C_p \,,V_G= C_g + C_g^1\, \phi \,,  \nonumber
\eeq
Now, the Friedmann equation reads, 
\beq
\label{ex1}
C_j (\dot{\phi}H)^2-C_p (\dot{\phi}H)^3-C_g^1 (\dot{\phi}H)+\frac{\Lambda}{3}=0 \,. \nonumber
\eeq
Note that $\dot{\phi}H$ appear as homogeneous powers of time in the algebraic equation (\ref{ex1}).
Hence since $H=1/t$ for Milne, taking,
\beq
\label{sol1}
\phi=\phi_0+\phi_1 t^2 \,,
\eeq
gives
\beq
12C_j (\phi_1)^2-24 C_p (\phi_1)^3-6C_g^1 (\phi_1)+\Lambda=0 \,.
\eeq
 Therefore we see that the integration constant $\phi_1$ is fixed by the cosmological constant for arbitrary values of the theory potentials $C_{Fab4}$. If the cosmological constant were to jump to a different value via, for example, an abrupt  phase transition then this would correspond to a change in $\phi_1$ i.e. to the scalar derivative being discontinuous. For self-tuning to work it is clear that we need at least one of 
$C_j, C_p$ or $C_g^1$ to be non zero. One can also check explicitly that the second FLRW equation is not independent and is obtained from the derivative of the Friedmann equation (along with its distributional part). All equations are therefore consistent for the self-tuning solution.

The above example illustrates in simple terms how self-tuning explicitly works in the Fab Four. To close off this section let us take a closer look at  what happens beyond the Fab Four. We will therefore consider for simplicity $V_G=0$, and take $F_J=c_J X^n$ and $F_P=c_P X^m$ where $m,n$ are some positive integers. The Friedmann equation reads,
\beq
\label{ex2}
(-1)^n \dot\phi^{2n+2} c_J H^2+(-1)^m c_P \dot\phi^{2m+3}H^3+\frac{\Lambda}{3}=0 \,,
\eeq
and it is in general an algebraic equation with respect to $\dot\phi$ which one can solve for arbitrary $\Lambda$. Setting $c_P$ or $c_J$ equal to zero one can find a simple solution of the type (\ref{sol1}) by arranging so as to get rid of time dependence in (\ref{ex2}). For example, taking $c_P=0$ we have the solution $\dot \phi=\phi_1t^{\frac{1}{n+1}}$ and hence (\ref{ex2}) simply gives,
\beq
(-1)^n \phi_1^{2n+2} c_J +\frac{\Lambda}{3}=0 \,,
\eeq
which gives the cosmological constant with respect to the integration constant $\phi_1$.

\section{Conclusions}

In this work, we have studied the extension of the Fab Four theories, introduced in \cite{Charmousis:2011bf}, in the context of the generalization of Horndeski theories presented in \cite{Gleyzes:2014dya}. In particular, we have shown that the two new Lagrangians considered in \cite{Gleyzes:2014dya} can be rewritten, up to some standard Horndeski terms, as very simple extensions of the John and Paul terms of the Fab Four, where the arbitrary functions of $\phi$ become arbitrary functions of $\phi$ and $X$. 
In summary, whereas the original Fab Four are characterized by four arbitrary functions of $\phi$ within the set of Horndeski theories that depend on four arbitrary functions of $\phi$ and $X$, the extended Fab Four are now characterized by two functions of $\phi$ and $X$ (for the extended John and Paul terms) and two functions of $\phi$ only (for the Ringo and George terms), within a set of theories that depend on six arbitrary functions of $\phi$ and $X$.

Another interesting application of the beyond Horndeski terms rewritten in the form
(\ref{bH4}) and (\ref{bH5}) are black hole hole solutions with non-trivial scalar
field configurations. It can be shown (see Appendix~\ref{a:bh} for details) that black hole solutions with non-trivial time-dependent scalar fields found 
in~\cite{Babichev:2013cya,Babichev:2015rva,Charmousis:2015txa}, can be extended to the beyond Horndeski theory.

Finally, let us comment about cosmology. 
 The Fab Four allow not only for a Minkowski metric with arbitrary cosmological constant but also for non trivial cosmological solutions, as required by construction. As shown in~\cite{Copeland:2012qf}, one can construct 
 cosmological solutions that mimick  radiation or matter eras by choosing appropriate functions for the Fab Four.  Ideally, one would like to recover a scenario close to  standard cosmology where the expansion is mostly driven by ordinary matter, with the scalar field screening the cosmological constant. It seems however that the scalar field generically tends to spoil this picture by giving a strong contribution to the cosmological evolution\footnote{Ed Copeland: private communication.}. By extending the Fab Four, one could hope that the extra freedom allowed by the dependence on $X$ could turn out to be helpful to get  more realistic scenarios. It would be interesting to explore this question in the future.

\acknowledgements
We  thank Ed Copeland for instructive discussions. 
This work was supported in part by National Science Foundation Grant No. PHYS-1066293 and
Russian Foundation for Basic Research Grants No. RFBR 13-02-00257, 15-02-05038, and
DL thanks the hospitality of the Aspen Center for Physics during the preparation of this work.
RS is supported by JSPS Postdoctoral Fellowships for Research Abroad.


\appendix

\section{The extended Fab Four actions in the unitary gauge}\label{a:unitarygauge}

In this section, we present the ADM formulation of extended Fab Four actions in the unitary gauge.

\subsection{The extended John and Paul actions in the unitary gauge}

We consider the ADM decomposition with respect to the uniform scalar field hypersurface $\Sigma_{\phi}$, which is orthogonal to the unit 4-vector,
	\begin{align}
		n_{\mu} \equiv -\frac{\partial_{\mu}\phi}{\sqrt{-X}} \,.
	\end{align}
We assume that $n_{\mu}$ is a time-like vector because $\phi=t$ in the unitary gauge.

In terms of the 4-vector $n_{\mu}$, the action (\ref{bH4}) can be rewritten as,
	\begin{align}
		S^{\rm ext}_J = \int\ d^4x\ \sqrt{-g}\  (-X)F_J(\phi,X)\ G^{\mu\nu} n_{\mu}n_{\nu} \,.
	\end{align}
Using the intrinsic and extrinsic 3d curvatures of the hypersurface $\Sigma_{\phi}$ respectively, $^{3}\!R_{\mu\nu}$ and $K_{\mu\nu}$, 
the Einstein tensor is decomposed as, 
	\begin{align}
		G^{\mu\nu} n_{\mu}n_{\nu} = \frac{1}{2}(^{3}\!R + K^2 - K^{\mu\nu}K_{\mu\nu}) \,.
	\end{align}
Here, $^{3}\!R$ and $K$ are their traces respectively.

Similarly, 
noting that the second derivative of $\phi$  in the action (\ref{bH5}) is projected onto the hypersurface $\Sigma_{\phi}$ because of the antisymmetry of $P^{\alpha\beta\mu\nu}$, 
the action (\ref{bH5}) can be rewritten as,
	\begin{align}
		S^{\rm ext}_P = -\int\ d^4x\ \sqrt{-g}\  (-X)^{\frac{3}{2}}F_P(\phi,X)\ P^{\alpha\beta\mu\nu} n_{\alpha}n_{\mu}K_{\beta\nu} \,,
	\end{align}
using the extrinsic curvature,
	\begin{align}
		K_{\beta\nu} \equiv (\delta^{\gamma}_{\beta} + n^{\gamma}n_{\beta})\nabla_{\gamma}n_{\nu} \,.
 	\end{align}
Using the Gauss-Codazzi equation, we find that the double dual of the Riemann tensor is decomposed as,
	\begin{align}
		P^{\alpha\beta\mu\nu} n_{\alpha}n_{\mu}K_{\beta\nu} = K^{\rho\sigma}\left( ^{3}\!R_{\rho\sigma} - \frac{1}{2}h_{\rho\sigma}{} ^{3}\!R \right) - \frac{1}{2}\left( K^3 -3K[K^2] + 2[K^3] \right) \,,
	\end{align}
where $h_{\mu\nu} \equiv g_{\mu\nu} + n_{\mu}n_{\nu}$, $[K^2] \equiv K^{\mu}_{\nu}K^{\nu}_{\mu}$, and $[K^3] \equiv K^{\mu}_{\nu}K^{\nu}_{\rho}K^{\rho}_{\mu}$.

Therefore, moving to the unitary gauge $\phi = t$, the actions (\ref{bH4}) and (\ref{bH5}) are presented in terms of the ADM variables as,
	\begin{align}
		S^{\rm ext}_J &= \int\ d^4x\ \sqrt{-g}\  \frac{(-X)F_J(\phi,X)}{2} \left( K^2 - K^{ij}K_{ij}  + ^{3}\!R \right)  \,, \\
		S^{\rm ext}_P &= \int\ d^4x\ \sqrt{-g}\  (-X)^{\frac{3}{2}}F_P(\phi,X)\ \left[  \frac{1}{2}\left( K^3 -3K[K^2] + 2[K^3] \right) - K^{ij}\left( ^{3}\!R_{ij} - \frac{1}{2}h_{ij}{} ^{3}\!R \right) \right]\,,
	\end{align}
with $X=-1/N^2$ where $N$ is the lapse function.
Comparing these results with eq. (3.3) in \cite{Gleyzes:2014dya}, 
we find that the Lagrangians for the extended John and Paul 
are equivalent to the beyond Horndeski Lagrangians ${\cal L}_4$ and ${\cal L}_5$ respectively with special relations
\footnote{With these expressions of $A_4$, $B_4$, $A_5$, and $B_5$ in terms of $F_J$ and $F_P$, we can also show that the covariant forms (3.9) and (3.10) in \cite{Gleyzes:2014dya} coincide with our expressions (\ref{bH4bis}) and (\ref{bH5bis}).},
	\begin{align}
		A_4 = B_4 = \frac{(-X)F_J}{2} \,,
	\end{align}
and
	\begin{align}
		A_5 = -\frac{B_5}{2} = \frac{(-X)^{\frac{3}{2}}F_P}{2} \,.
	\end{align}

\subsection{About the extension of George and Ringo}
We have seen that the additional Lagrangians in theories beyond Horndeski 
can be rewritten, up to Horndeski terms, as  extensions of John and Paul. By contrast, one finds that  the analogous extensions of George and Ringo,
	\begin{align}
		S^{\rm ext}_G &\equiv \int {\rm d}^4 x \sqrt{-g} \, F_G(\phi,X)\, R\,, \label{eG} \\
		S^{\rm ext}_R &\equiv \int {\rm d}^4 x \sqrt{-g}\, F_R(\phi,X)\, \hat{G} \,, \label{eR}
	\end{align}
		lead to ADM Lagrangians that  depend on the time derivative of the lapse function, $\dot{N}$.
This is because, in contrast to the Einstein tensor and the double dual of the Riemann tensor, the scalar curvature and the Gauss-Bonnet term contain terms with the normal derivative of the extrinsic curvature, which correspond to the Gibbons-Hawking boundary term and its generalization \cite{Myers:1987yn}.
Integrating by parts, they give the following terms in the actions,
	\begin{align}
		S^{\rm ext}_G &\supset \int {\rm d}^4 x\ \sqrt{-g}\ (n^{\mu}\nabla_{\mu}F_G)K \,, \label{GHeG} \\
		S^{\rm ext}_R &\supset -\int {\rm d}^4 x\ \sqrt{-g}\ (n^{\mu}\nabla_{\mu}F_R)\left[ 8K^{\rho\sigma}\left( ^{3}\!R_{\rho\sigma} - \frac{1}{2}h_{\rho\sigma}{} ^{3}\!R \right) + \frac{4}{3}\left( K^3 -3K[K^2] + 2[K^3] \right) \right] \,. \label{GHeR}
	\end{align}
When $F_G$ or $F_R$ only depend on $\phi$, these terms correspond to the beyond Horndeski Lagrangians ${\cal L}_3$ and ${\cal L}_5$ respectively in the unitary gauge. 
On the other hand, when $F_G$ or $F_R$ depend on $X$, they give terms linear in $\dot{N}$ because $X=-1/N^2$ in the unitary gauge.  
Therefore, these Lagrangians do not belong to the simple class of ADM Lagrangians studied in \cite{Gleyzes:2014dya,Gleyzes:2014qga} and the Hamiltonian analysis given in those references. 


\section{Black holes in beyond Horndeski theory}\label{a:bh}

The way we have rewritten the beyond Horndeski terms  (\ref{bH4}) and (\ref{bH5}) turns out to be extremely convenient to 
find black hole solutions with non-trivial non-trivial scalar field configurations. 
It has been shown in a number of papers that the John term leads to regular black hole solutions with a time-dependent galileon
field~\cite{Babichev:2013cya,Babichev:2015rva,Charmousis:2015txa}. 
In particular, there exists a stealth black hole configuration with a non-trivial
time-dependent scalar field and a Schwarzschild metric geometry for spacetime 
(hence the name stealth, due to the fact that the scalar field does not affect the
metric for this solution). 
An important property of this solution is  that $X=$const. Using this property and the
equations of motion 
$${\cal E}^{\mu\nu}_J =0, \,\,  {\cal E}^{\phi}_J =0,$$ 
with 
${\cal E}^{\mu\nu}_J$ and ${\cal E}^{\phi}_J $ given by (\ref{eomJg}) and
(\ref{eomphi}) correspondingly, 
it is not difficult to see that the stealth solution found in
\cite{Babichev:2013cya} remains a solution for the extended John term. 
Indeed, the scalar field equation ${\cal E}^{\phi}_J =0$ is automatically satisfied
because the Einstein tensor vanishes for the Schwarzschild metric, while 
the metric equations ${\cal E}^{\mu\nu}_J =0$ reduce to the metric equations for
pure John, up to the overall factor $F_J$, since the last term of 
(\ref{eomJg}) is zero because $G_{\alpha\beta}=0$ and the other terms containing
derivatives of $F_J$ vanish thanks to the property $X={\rm const}$.
In the context of the extended John term, it would be interesting 
to investigate the existence and properties of the counterparts of other
(non-stealth) solutions, found in
\cite{Babichev:2013cya,Babichev:2015rva,Charmousis:2015txa,Charmousis:2015aya}.


\end{document}